\documentclass[useAMS,usenatbib, usegraphicx]{mn2e}
\usepackage{float}
\usepackage{graphicx}
\usepackage{amssymb}
\usepackage{amsmath}
\usepackage{cleveref}
\usepackage[english]{babel}
\usepackage{multirow}
\usepackage[T1]{fontenc}
\usepackage{aecompl}

\def\apj{ApJ}
\def\mnras{MNRAS}
\def\nat{Nat}

\def\araa{ARA\&A}                % "Ann. Rev. Astron. Astrophys."
\def\aap{A\&A}                   % "Astron. Astrophys."
\def\apjl{ApJ}                   % letter at ApJ
\def\pasj{PASJ}

\begin{document}

\title{Spectral State Transitions of the Ultraluminous X-ray Source IC 342 X-1}

\author[H. Marlowe et al.]
{\parbox[]{6.in}{H.~Marlowe,$^1$ P.~Kaaret,$^1$ C.~Lang,$^1$ H.~Feng,$^2$ F.~Gris\'{e},$^{3,4,5}$  N.~Miller,$^6$ D.~Cseh,$^7$ S.~Corbel,$^8$ R.~F.~Mushotzky$^9$}\\\\
\footnotesize
$^1$Department of Physics and Astronomy, University of Iowa, Van Allen Hall, Iowa City, IA 52242, USA; \\hannah-marlowe@uiowa.edu\\
$^2$Department of Engineering Physics and Center for Astrophysics, Tsinghua University, Beijing 100084, China\\
$^3$Instituto de Astrof\'{i}sica de Canarias, E-38200 La Laguna, Tenerife, Spain\\
$^4$Departamento de Astrof\'{i}sica, Universidad de La Laguna, Avda. Astrof\'{i}sico Francisco Sanchez s/n, E-38271 La Laguna, Tenerife, Spain\\
$^5$Observatoire astronomique de Strasbourg, Universit\'{e} de Strasbourg, CNRS, UMR 7550, 11 rue de l'Universit\'{e}, F-67000 Strasbourg\\
$^6$Department of Mathematics and Physical Sciences, Stevenson University, Stevenson, MD 21153\\
$^7$Department of Astrophysics/IMAPP, Radboud University Nijmegen, P.O. Box 9010, 6500 GL Nijmegen, The Netherlands\\
$^8$Universit\'{e} Paris 7 Denis Diderot and Service d'Astrophysique, UMR AIM, CEA Saclay, F-91191 Gif sur Yvette, France\\
$^9$Department of Astronomy, University of Maryland, College Park, MD 20742, USA\\
}
\maketitle

\begin{abstract}
We observed the Ultraluminous X-ray Source IC 342 X-1 simultaneously in X-ray and radio with \textit{Chandra} and the \textit{JVLA} to investigate previously reported unresolved radio emission coincident with the ULX. The \textit{Chandra} data reveal a spectrum that is much softer than observed previously and is well modelled by a thermal accretion disc spectrum. No significant radio emission above the rms noise level was observed within the region of the ULX, consistent with the interpretation as a thermal state though other states cannot be entirely ruled out with the current data. We estimate the mass of the black hole using the modelled inner disc temperature to be $30~\mathrm{M_{\odot}} \lesssim M\sqrt{\mathrm{cos}i}\lesssim200~\mathrm{M_{\odot}}$ based on a Shakura-Sunyaev disc model. Through a study of the hardness and high-energy curvature of available X-ray observations, we find that the accretion state of X-1 is not determined by luminosity alone. 

\end{abstract}
%\doublespace
\begin{keywords}
X-rays: binaries  -- accretion, accretion discs -- black hole physics-- radio continuum: general.
\end{keywords}

%Introduction

\section{Introduction}

Ultraluminous X-ray sources (ULXs) are non-nuclear, extragalactic accreting compact objects whose X-ray luminosities exceed the Eddington limit for stellar mass black hole binaries (BHB) with mass $\lesssim$~20~M$_{\rm \odot}$. Their high luminosities suggest they are either intermediate mass black holes (IMBH) \citep{Colbert99, Makishima00, Kaaret01}, that their emission is beamed \citep[e.g.][]{Poutanen07,King09}, or that they are stellar-mass BHBs that are emitting at super-Eddington rates \citep[e.g.][]{Begelman02}. Beaming alone cannot account for the excess luminosity from all sources, as demonstrated by observations using He II recombination lines in surrounding nebula for the ULX Holmberg II X-1 and for NGC 5408 X-1 showing the observed luminosity to be consistent with relatively isotropic emission \citep{Pakull02, Kaaret04, Kaaret09a}. Likely the ULX population is explained by some combination of these models. By studying the unique characteristics and spectral evolution of ULXs, and comparing their spectral and timing features to the well known properties of BHBs, we can hope to elucidate the physical processes that drive these sources. 
 
Spectral curvature above $\sim$6~keV in the hard component of a large fraction of ULX spectra \citep{Stobbart06, Gladstone09, Miyawaki09} distinguishes them from BHBs, whose hard component is generally modelled by a powerlaw which extends unbroken to much higher energies, often above $\sim$100~keV. With sensitivity above 10~keV, recent observations with the Nuclear Spectroscopic Telescope Array (\textit{NuStar}) have confirmed a spectral cutoff in ULXs such as NGC 1313 X-1 \citep{Bachetti13}. \citet{Gladstone09} suggested that the high energy roll-over and soft spectral feature seen in many ULX spectra may correspond to a super-Eddington state of accretion onto a stellar-mass BH (the `ultraluminous' state). The spectral curvature is thought to be due to heavy Comptonization of disc photons by a cool, optically-thick corona. Further study of ULX populations has motivated the classification of three distinct ULX states: a \textit{Broadened disc} state, and two-component \textit{hard ultraluminous} and \textit{soft ultraluminous} states \citep{Sutton13}. The broadened disc state is thought to occur near the Eddington accretion rate, while the latter two likely represent super-Eddington accretion states.

For a single source, transitions between accretion states in the ultraluminous classification scheme are ascribed solely to changes in the accretion rate. Inclination is also expected to affect the state classification \citep{Poutanen07, Sutton13}, but would not change over time for a single source with a fixed orientation to the observer. Luminosity is often used as a predictor of accretion rate, and in BHBs it is well established that the spectral state is not set by the luminosity alone \citep[for a review, see][]{Remillard06}. \citet{Grise10} and \citet{Vierdayanti10} have shown that the spectral state varies at a single luminosity for the ULXs Holmberg II X-1 and Holmberg IX X-1, respectively, suggesting that an additional parameter beyond accretion rate can vary for individual sources. It would be interesting to discover whether this trait is common to other ULXs and, thus, an additional parameter is needed to set the accretion state of a given source.

IC 342 is a nearby (3.9 Mpc; \citealt{Tikhonov10}) starburst galaxy that contains two ULXs, X-1 and X-2. X-1 is surrounded by a large optical and radio nebula that may be an X-ray ionized bubble driven by strong outflows from the ULX \citep{Pakull02, Roberts03, Feng08}. X-1 and X-2 were the first ULXs reported to transition between distinct spectral states \citep{Kubota01}. X-1 transitioned from an apparently disc-dominant state to a powerlaw-dominant state while the spectrum of X-2 coincidentally made the reverse transition. 

The X-ray spectrum of IC 342 X-1, hereafter X-1, has previously been observed in a spectral state characterized by a high luminosity, apparently cool disc, and a spectral turnover at high energy \citep{Feng09, Yoshida12}. \citet{Feng09} reported an anti-correlation between the X-ray luminosity and temperature of the soft energy feature of X-1, ruling out interpretation of the soft component as thin disc emission. A negative temperature-luminosity correlation is predicted if the soft component is caused by optically thick outflows thought to occur in a super-Eddington accretion state \citep{Poutanen07, Kajava09}.  Therefore, X-1 is a strong candidate for a source which enters super-Eddington or 'ultraluminous' accretion states.

In \textit{VLA} observations of X-1 in 2007 and 2008, \citet{Cseh11, Cseh12} reported possible unresolved radio emission coincident with the ULX which was not detected with follow-up \textit{EVN} observations. To confirm or deny the existence of a self-absorbed compact jet and investigate whether a relationship known as the fundamental plane (FP) of black hole activity \citep{Merloni03,Falcke04} could be used to estimate the mass of the BH, we obtained simultaneous radio and X-ray observations of X-1.

In this paper, we describe the spectral properties of the recent \textit{Chandra} X-ray and \textit{JVLA} radio observations of X-1. We also analyse previous high quality X-ray observations of the source in a systematic fashion to investigate the evolution of the system over time via its X-ray intensity, spectral hardness, and high-energy curvature. We additionally estimate the mass of the BH using fit parameters from the most recent \textit{Chandra} observation.

\section{X-ray Observations and Reduction}

We analysed a new \textit{Chandra} observation of X-1 and completed analysis of the previous high-quality observations of X-1 in order to examine the evolution of its spectral properties. The data were reduced using \textsc{heasoft} (Version 6.13) with up-to-date calibration files as of May 2013. Reduction and analysis details for each instrument are described in this section.

The observations, with references to previous analysis in the literature and observation identifiers used in this paper, are summarized in Table \ref{table:prev}. We have chosen to include \textit{Chandra} observations described by \citet{Roberts04} only in the hardness-intensity diagram due to pileup concerns. Using the \textsc{xspec} tool \textit{pileup map}, we find that these observations have maximum count-rates per frame of \textgreater~0.14 in the region of X-1, corresponding to a pileup fraction greater than 5~per~cent. We exclude all available  \textit{Swift}  observations because they have too few counts to confidently constrain the spectral properties.

\begin{table*}
\centering
\caption{Summary of Observations}
%\tabcolsep=0.01cm
%\centering
\small
\begin{tabular}{ccccccc} 
\hline
Identifier & Date & Instrument & Exposure & Counts & ObsID & Ref\\
\hline
(1) & (2) & (3) & (4) & (5) & (6) & (7)\\
\hline
ASCA1 & 19/09/1993 & \textit{ASCA} & 77.7 & 16621 & 60003000 & A, B, C, D, E \\%\citep{Okada98,Makishima00, Kubota01, Mizuno01,Watarai01}\\ 
ASCA2 & 24/02/2000 & \textit{ASCA} & 545 & 15804 & 68001000   & C, D, E \\%\citep{Kubota01, Mizuno01, Watarai01}\\ 
XM1 & 11/02/2001 & \textit{XMM-Newton} & 9.5  & 3932 & 0093640901 & F, G, I, J, K, L \\ %\citep{Feng09,Mak11, Yoshida12,Sutton13, Pintore14}\\
XM2 & 20/02/2004 & \textit{XMM-Newton} & 21.2 & 22962 & 0206890101 & G, I, J, L \\%\citep{Feng09, Mak11,Yoshida12,Pintore14}\\
XM3 & 17/08/2004 & \textit{XMM-Newton} & 23.4 & 13026 & 0206890201 & G, H, I, J, L \\% \citep{Feng09,Gladstone09,Mak11,Yoshida12,Pintore14}\\
XM4 & 10/02/2005 & \textit{XMM-Newton} & 7.6 & 12943 & 0206890401 & G, I, J, L \\%\citep{Feng09, Mak11,Yoshida12,Pintore14}\\
SU1 & 07/08/2010 & \textit{Suzaku} & 74.4 & 22059 & 705009010 & J \\ %\citep{Yoshida12}\\
SU2 & 20/03/2011 & \textit{Suzaku} & 75.5 & 23408 & 705009020 & J \\ %\citep{Yoshida12}\\
CHR & 26/08/2002 & \textit{Chandra} & 9.9 & 2313 & 2917 & J\\
CH & 29/10/2012 & \textit{Chandra} & 9.1 & 2245 & 13686 & This Paper\\

\hline
\end{tabular}
\begin{flushleft}
(1) Identifiers follow convention of \citet{Yoshida12}; (2) Date of observation; (3) Instrument; (4) Duration of observation after filtering (ks); (5) Total counts in spectrum included in fitting; (6) Observation ID; (7) Selected reference to previous analysis in literature.\\
(A) \citet{Okada98}; (B) \citet{Makishima00}; (C) \citet{ Mizuno01}; (D) \citet{Watarai01}; (E) \citet{Kubota01}; (F) \citet{Wang04}; (G)  \citet{Feng09}; (H) \citet{Gladstone09}; (I) \citet{Mak11}; (J) \citet{Yoshida12}; (K) \citet{Sutton13}; (L) \citet{Pintore14}.\\
\end{flushleft}
\label{table:prev}
\end{table*}

\subsection{New \textit{Chandra} Observations}
X-1 was observed with \textit{Chandra}'s Advanced CCD Imaging Spectrometer (ACIS) for 9.13 ks in its FAINT mode on October 29 2012 beginning at 07:10:25~UTC (ObsId 13686), coincident with the \textit{JVLA} radio observations. X-1 is a bright source with an ACIS counting rate $\sim0.2~c~s^{-1}$ and has suffered from pileup in previous \textit{Chandra} observations. To avoid pileup, the observations were carried out using offset pointing ($3.43'$), operating only the S3 chip and using a 1/8 sub-array to reduce the frame time to 0.4 sec. Data reduction was performed using the \textit{Chandra} Interactive Analysis of Observation software package (\textsc{ciao}) (data processing version 8.4.5, \textsc{caldb} version 4.5.2), beginning with the level 2 event files.  \textit{dmextract} was used to create a light curve of the events on the S3 chip to look for background flares. No flares were observed and the maximum count rate was below 2.5 counts per second. 

Offset pointing increases the positional uncertainty approximately to 1/4 times the 50~per~cent encircled energy radius at the offset position\footnote{http://cxc.harvard.edu/cal/ASPECT/celmon/}. For our $3.43'$ offset and a photon energy of 5 keV, we used the \textsc{ciao} utility \textit{mkpsfmap} to find the encircled energy radius and determined a positional uncertainty of $0.33"$. The \textsc{ciao} \textit{wavdetect} utility was used to search for point sources in the field using a significance threshold of $10^{-6}$ and returned 4 sources. The ULX is identified with a source centred at R.A. = $3^{h}45^{m}55.55^{s}$, decl. = $+68^{\circ}04'55.25''$ (J2000.0), consistent with the previously determined location of the ULX by \citet{Feng08} within positional uncertainties.

The \textsc{ciao} \textit{specextract} utility was used to extract a spectrum and create unweighted, point-source aperture corrected ARF and RMF response matrices for events on the S3 chip using a 4" circular region centred at R.A. = 3:45:55.594, decl. = +68:04:55.39, enclosing $\sim$90~per~cent of the PSF at 8 keV. The background is specified by a surrounding annular region with an inner radius of 15.17" and outer radius of 28.7". \textsc{xspec} version 12.8.0 \citep{Arnaud96} was used to fit the spectrum in the 0.7 -- 8~keV band. Events were grouped to 15 counts per bin.

\subsection{ \textit{ASCA} }\label{sec:asca}
We screened event files for two \textit{ASCA} observations using the \textsc{heasoft} \textsc{ascascreen} tool with standard REV2 screening procedures.  A circular region of 3 arcmin radius was used to extract the spectrum of X-1 in both of the \textit{ASCA} observations, encircling more than 60~per~cent of the energy at 4.5 keV (SIS) and avoiding overlap with IC 342 X-2. There are three additional X-ray sources which overlap this region, sources 6, 12 and 9 which are described by \citet{Kong03}. We mask out sources 6 and 12 using circular regions of 77 arcsec radius. Source 9 is unresolved by \textit{ASCA}, but is not expected to cause significant contamination as its flux is approximately 1/10 that of X-1 in the 0.2--12--keV band (and only 4 percent for the brighter ASCA1 observation). The luminosity of source 9 also appears constant over several years within errors \citep{Kong03, Mak08}. The inferred fluxes do not depend on the extraction region size as corrections for encircled energy fraction are included in the response files. The \textsc{heasarc} \textit{Ftools} interface, \textsc{xselect}, was used to extract spectra for both GIS and SIS instruments. An energy range of 0.7--10 keV was used to fit the spectra due to the cutoff in instrumental sensitivity outside of this range.

Both the SIS and GIS instruments were analysed from the 1993 \textit{ASCA} (ASCA1) observation. Because the source falls across two chips of the SIS detector, the response files were created for the source taking into account the gap between chips. \textit{sisrmg} was used to create the SIS response matrices for each side of the gap, which were combined using \textsc{addrmf} with weighting by total counts within the source region on each CCD chip. SIS0 and SIS1 spectra were combined using \textsc{addascaspec} for bright and bright2 mode data separately. The SIS and GIS spectra were fit simultaneously, and a constant multiplier was fit in front of the GIS spectrum to account for its lower response compared to the SIS instrument; the constant was found to range between 1.1-1.15. Background files were created for the GIS using blank sky data with the tool \textit{mkgisbgd}, and the GIS2 and GIS3 spectra and ancillary response file (arf) were combined using the tool \textit{addascaspec}. For the ASCA2 observation, we consider only GIS instrument spectra due to degradation of the SIS instrument over time. Final spectra were grouped to a minimum of 60 counts per bin using the tool \textit{grppha} before fitting. \\

\subsection{ \textit{XMM-Newton} }
We created event files for four  \textit{XMM-Newton}  observations of X-1 using \textsc{SAS} 12.0.1 with up-to-date calibration files. In all observations, the source is specified by a circular 30~arcsec radius extraction region and a background region is located nearby on the same chip as the source. The same background region was used for all four  \textit{XMM-Newton}  observations. The reduction otherwise follows the same procedure as outlined in \citet{Feng09}. The MOS and PN spectra were fit simultaneously, and a constant, which ranged between 1.0--1.1, was fit in front of the MOS spectra to account for their lower response compared to the PN instrument. Events were grouped to a minimum of 25 counts per bin and an energy range of 0.4--10 keV was used to fit the spectra due to the cutoff in instrumental sensitivity outside of this range.

\subsection{\textit{Suzaku}}
We extracted spectra and created response files from XRT data for two \textit{Suzaku} observation of X-1 using \textsc{xselect}. The data from the two remaining front illuminated (FI) and one back-illuminated (BI) XIS CCDs were reprocessed through the \textit{aepipeline}, a third FI CCD (XIS2) was catastrophically damaged in 2006 November and has ceased useful operation. Events were screened out during and 436 seconds following passage through the SAA, and only events where pointing was $\geq 5^{\circ}$ above Earth and $\geq 20^{\circ}$ above the sunlit limb of Earth were accepted.  The source was extracted from a circular region of 140~arcsec radius, encircling more than 60~per~cent of the energy at 4.5~keV (XRT) with circular 1~arcmin radius masks used to exclude the same two weak sources as for \textit{ASCA}. The weak source 9 is also unresolved in the \textit{Suzaku} data as it is for \textit{ASCA}, but again contributes approximately 1/10 times the flux of X-1. A circular region of 140~arcsec was used to create background spectra for both observations. Inferred fluxes are independent of the region size as the response files provide correction for the encircled energy fraction. Due to their very similar responses, spectra from the two operational FI CCDs were combined. The FI and BI spectra were each grouped to a minimum of 60 counts per bin using the tool \textit{grppha}. An energy range of 0.7--10~keV was used to fit the spectra due to the cutoff in instrumental sensitivity outside of this range. We fit the FI and BI spectra simultaneously, allowing a constant multiplier ranging between 1.0--1.1 in front of the BI spectrum to account for the different normalizations of the instruments. Following \citet{Yoshida12}, we exclude the region of 1.8--2.0~keV due to calibration uncertainties.

\section{Spectral Analysis}

We used \textsc{xspec} (version 12.8.0) to fit the spectra. Fluxes and luminosities were calculated using the \textsc{xspec} model \textit{cflux} assuming a distance of 3.9 Mpc and all errors are given to the 90~per~cent confidence level.

\subsection{Hardness Evolution}
We investigated the spectral hardness of X-1 over all epochs in our sample by creating a hardness-intensity diagram (HID). \citet{Pintore14} carried out color-color analysis of several ULXs and found that sources grouped into two clusters, which seem to correspond to sub and super-Eddington accretion states. Their analysis used only \textit{XMM-Newton} EPIC-pn data and so took colors as a ratio of counts in each energy band. Because our sample includes observations from many instruments, we cannot simply calculate the hardness using relative counts due to cross calibration uncertainties. Instead, we calculated the unabsorbed flux in each band. This is similar to the procedure of \citet{Sutton13} who fit spectra with a Comptonization plus MCD model and calculated the intrinsic fluxes in two energy bands. To minimize the model dependency of our flux calculation, we fitted each observation in our sample with an absorbed powerlaw model using a fitting range restricted to the hard and soft bands separately. The absorption value was fixed to a value of $0.723\mathrm{\times10^{22}cm^{-2}}$, found by fitting all the spectra simultaneously over the entire 0.7--8.0~keV range.

We calculated the flux in a hard energy band (4--8~keV) and a soft energy band (0.7--4.0~keV), these are comparable to bands 2+3 and 4 in \citet{Pintore14}, and compared their ratio to the total flux in the 0.7--8.0~keV range. Calculating the intrinsic flux in each band allows the absorption to be taken into account, which is a significant concern for X-1. \citet{Pintore14} showed that the high absorption column of X-1 prevented the source from falling into the same color-color groups as other ULXs with similar spectra when the soft band included counts from 0.7--2.0~keV. Using the modelled intrinsic fluxes, we find that the HID looks very similar using a soft band of either 2.0--4.0~keV or 0.7--4.0~keV, thus we argue that this is likely a better method for sources with large absorption columns.

\begin{figure}
        \centering

               \includegraphics[width=.48\textwidth]{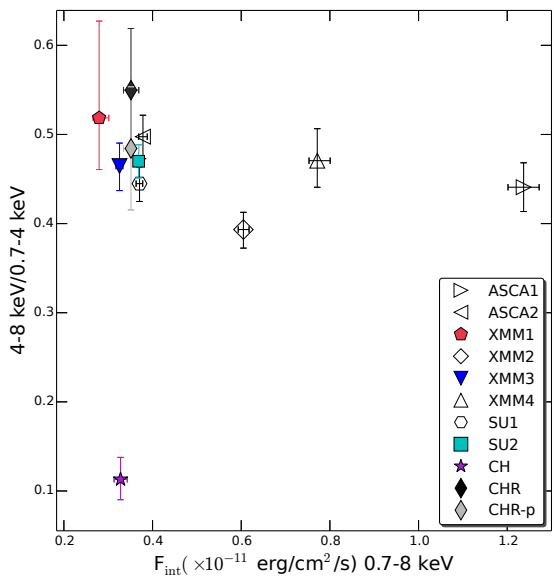} \\
             
                \caption{Ratio of the model flux in the 4.0--8.0~keV range to the 0.7--4.0~keV range versus total intrinsic flux, in units of $\rm erg~s^{-1}~cm^{-2}$, from 0.7--8.0~keV. Hollow symbols mark observations classified as ultraluminous based on the apparent high energy curvature of the spectra as described in Section 3.2. The point labelled CH is our new \textit{Chandra} observation, whereas the point CHR represents the previous \textit{Chandra} observation by \citet{Roberts04}. A gray diamond labelled CHR-p denotes the hardness ratio of CHR after a 10 per cent pileup correction is applied.}
\label{fig:hardness}                 
\end{figure}

The resulting hardness ratios versus total intrinsic flux are plotted in Figure \ref{fig:hardness}. We find that the \textit{Chandra} observation, filled star labelled CH, is significantly softer that any other observation in our sample. To test whether the softness of the \textit{Chandra} observation might be due to instrumental sensitivity, we re-analysed a 9.9~ks observation of X-1 made with \textit{Chandra} on August 26, 2002 (sequence number 600254) when the source was observed to be in a powerlaw-dominant state \citep{Roberts04}. These data, hereafter CHR, were reduced following the same procedure as for the new \textit{Chandra} observation. We find that the previous \textit{Chandra} observation, marked by a filled diamond labelled CHR, groups with the harder/low luminosity observations in our sample. However, this observation suffers from mild pileup of 5--10 per cent, which could potentially cause the spectrum to appear artificially hardened. We therefore estimated a corrected hardness of the spectrum by moving 10 per cent of events from the hard to soft energy bands, simulating the worst-case artificial hardening, and find that the HR falls within the uncertainty of the uncorrected value. The pileup-corrected point is included in Figure \ref{fig:hardness} as a filled gray diamond. Thus, we conclude that the apparent softness of the recent \textit{Chandra} observation is real and not an instrumental effect.

The variation in the hardness ratio over a narrow flux band is significant. Fitting a constant to the observations with fluxes between 3--4$\mathrm{\times10^{-12}erg~s^{-1}~cm^{-2}}$ gives $\chi^{2}$ = 206.61 for 6 DoF, showing that the hardness ratios are strongly inconsistent with a constant value. We additionally calculate the deviation of the \textit{Chandra} observation from the weighted average of the other observations and find the difference to be 13.3 times the 90 per cent error. Therefore, the source's spectral evolution clearly does not depend on the luminosity alone. \\

\subsection{High Energy Curvature and the Ultraluminous State}

To investigate the existence of a high energy roll-over in the observations of X-1, we follow the procedure of \citet{Gladstone09} and fit the spectra with a broken powerlaw (\textit{bknpower} in \textsc{xspec}) in the 2--10~keV range and compared the goodness of fit to single component models in the same range. Because the single component absorption columns are generally constrained to be below $\sim 1 \times 10^{22}\ \mathrm{cm^{-2}}$ for X-1, the absorption has little effect on data above 2~keV and is poorly constrained in this energy range. We therefore do not include an absorption component in the fits. 

\citet{Gladstone09} suggest that the spectral curvature of the ultraluminous state can only be reliably characterized in high quality observations with \textgreater 10,000 counts. Three of the nine observations in our study have fewer than 10,000 counts. The recent \textit{Chandra} observation is excluded from the table due to its low counts and the low sensitivity of the instrument above 8 keV. We also exclude XMM1 due to its low total counts, though we include XMM4 which has more than 8000 counts. The broken powerlaw fit parameters are listed in Table \ref{table:breakfit}. For the expected spectral curvature, we require that $\Gamma_{2}$\textgreater$\Gamma_{1}$ and E$_{\rm Break}$ be above 2~keV.

\begin{table*}
\caption{Power-law vs. Broken Power-law}
\centering
%\centering

%\begin{tabular}{l*{7}{c}r}
%\hline
% $N_{H}$ & $\Gamma$ & $N_{PL}$ & $T_{in}$ & $f_{X}$  & $L_{X}$ & $\chi^{2}/dof$  \\
%\tabcolsep=0.1cm

\begin{tabular}{ccccccccc}  %{l*{10}{c}r}

\hline
 &Power-law &&Broken Power-law&&&&&\\
Identifier & $\Gamma$ & $\chi^2$(DoF) & $\Gamma_{1}$  & $E_{Break}$ & $\Gamma_{2}$ & $\chi^2$(DoF) & $\Delta\chi^{2}$ & P(F-Test)\\
%     	&	    ($10^{22}\ \mathrm{cm^{-2}}$) & & $\times10^{-4}$     &        & ($\times 10^{-12}\ \mathrm{erg\ s^{-1}\ cm^{-2}}$) & ($\times 10^{39}\ \mathrm{erg\ s^{-1}}$) & ($\times 10^{-12}\ \mathrm{erg\ s^{-1}\ cm^{-2}}$) &\\
\hline
(1) & (2) & (3) & (4) & (5) & (6) & (7) & (8) & (9)\\
\hline
%\multicolumn{9}{c}{Power-law} \\

%XMM1 & Pow &  $1.06^{+0.10}_{-0.09}$ & $2.94^{+0.14}_{-0.13}$ & $2.3_{-0.3}^{+0.4}$ & ... & ... & ... & ... &... & $7.2^{+1.0}_{-0.9}$ & $13.1^{+1.8}_{-1.7}$ & $7.2^{+1.0}_{-0.9}$ & ... & 141.48(110) \\

%XMM1 & Cutoff-Pow &  $0.53^{+0.15}_{-0.15}$ & $0.0^{+0.8}_{-0.8}$ & $1.7_{-0.3}^{+0.3}$ & ... & $1.07_{-0.3}^{+0.4}$ & ... & ... &... & $2.9^{+0.5}_{-0.4}$ & $5.2^{+0.9}_{-0.8}$ & ... & ... & 99.48(108) \\

%\hline
%\multicolumn{9}{c}{Power-law + disc blackbody} \\
%XMM1 & Diskbb & $0.56^{+0.06}_{-0.05}$ & ...  & ... & $0.91^{+0.05}_{-0.05}$ & ... & ... & $0.23^{+0.05}_{-0.05}$ &... & $3.01^{+0.16}_{-0.16}$ & $5.5^{+0.3}_{-0.3}$  & ... & $3.01^{+0.16}_{-0.16}$ & 101.92(110)   \\

%diskbb compare ASCA1 & - & 121.32(138) & $1.09^{+0.15}_{-0.17}$  & $3.4^{+0.3}_{-0.3}$ & $2.13^{+0.14}_{-0.12}$ & 117.14(136) & 4.18   & 0.09 \\%***
ASCA1 & $1.66^{+0.04}_{-0.04}$ & 203.53(138) & $1.09^{+0.15}_{-0.17}$  & $3.4^{+0.3}_{-0.3}$ & $2.13^{+0.14}_{-0.12}$ & 117.14(136) & 86.39   & 5e-17 \\%***

ASCA2 & $1.52^{+0.05}_{-0.05}$ & 117.10(133) & $1.32^{+0.12}_{-0.14}$  & $3.9^{+2.3}_{-0.6}$ & $1.78^{+0.75}_{-0.13}$ & 102.93(131) & 14.17   & 2e-4 \\

%XMM1 & $1.83^{+0.04}_{-0.04}$ & 510.89(380) & $1.37^{+0.10}_{-0.11}$  & $3.66^{+0.18}_{-0.18}$ & $2.30^{+0.11}_{-0.10}$ & 420.01(378) & 90.88   & 1-8.51835e-17 \\ not enough counts

XMM2 & $1.81^{+0.04}_{-0.04}$ & 455.94(379) & $1.38^{+0.09}_{-0.12}$  & $3.69^{+0.17}_{-0.21}$ & $2.26^{+0.10}_{-0.11}$ & 375.49(377) & 80.45   & 1.3e-16 \\%***

XMM3 & $1.57^{+0.06}_{-0.06}$ & 222.14(226) & $1.52^{+0.07}_{-0.03}$  & $6.6^{+0.8}_{-1.4}$ & $2.4^{+1.0}_{-0.7}$ & 215.86(224) & 6.28   & 0.04 \\%***

XMM4 & $1.67^{+0.05}_{-0.05}$ & 250.59(212) & $1.29^{+0.13}_{-0.13}$  & $4.0^{+0.4}_{-0.3}$ & $2.22^{+0.21}_{-0.17}$ & 205.61(210) & 44.98   & 1.0e-9\\%***

Su1 & $1.68^{+0.04}_{-0.04}$ & 206.61(195) & $1.54^{+0.07}_{-0.07}$  & $5.1^{+0.4}_{-0.4}$ & $2.3^{+0.3}_{-0.2}$ & 180.64(193) & 25.97   & 2e-6 \\ %***

Su2 & $1.51^{+0.04}_{-0.04}$ & 232.30(207) & $1.33^{+0.10}_{-0.34}$  & $3.9^{+0.9}_{-1.1}$ & $1.70^{+0.12}_{-0.13}$ & 218.93(205) & 13.37   & 0.02\\

\hline

%[get errors for the discbb and pow fits, make sure they are consistent with how the combo fit was produced]
\end{tabular}
\label{table:breakfit}
%\end{center}
%

\begin{flushleft}
(1) Observation identifier; (2) Photon Index of PL Model; (3) Chi-Squared and degrees of freedom for PL fit; (4) Photon index (Bknpower); (5) Break Energy (keV); (6) Photon index past break; (7) Chi-Squared and degrees of freedom (Bknpower); (8) Chi-square power-law - Broken Power-law; (9) F-test false rejection probability
\end{flushleft}

\end{table*}

Of the seven observations considered, five are improved with statistical probability of \textgreater 99~per~cent using the F-test.  XMM3 is the least improved, with probability 96~per~cent. We note that \citet{Gladstone09} performs similar analysis of the same XMM3 observation and quote an F-test probability of \textgreater 99~per~cent when compared to the powerlaw model. However, this seems to be a typographical error, and their fit is actually only preferred to 97~per~cent based on their tabulated $\chi^{2}$ values. 

For comparison we carried out a fit of the spectra of XMM1 and CH despite their low counts, using 0.4--10.0~keV (XMM1) and 0.7--8.0~keV (CH) bands with an absorption factor to utilize as many counts as possible. We find that the photon index breaks to a shallower value in the case of XMM1, disqualifying it based on the criteria above. The broken powerlaw fit of CH is preferred \textgreater 99~per~cent when compared to a powerlaw fit. However, CH is better fit by a thermal disc spectrum than a powerlaw ($\chi^2_{disk}$ = 98.99(108); $\chi^2_{pow}$ = 138.70(108)), which causes an exponential cutoff in the spectrum at high energies. Therefore, a broken powerlaw is expected to provide significant improvement over a powerlaw in this case.

We classify into the ultraluminous state those observations where a broken powerlaw fit is preferred at \textgreater 99 per cent confidence (excepting the soft, disc CH spectrum) to a simple powerlaw. These observations are represented by hollow symbols in Figure \ref{fig:hardness}. We note that \citet{Rana14} have recently confirmed spectral curvature in X-1 up to 30~keV with \textit{XMM-Newton} and \textit{NuStar} observations. Using their cutoff powerlaw fit parameters, we find that this observation would fall in the hard spectral track in Figure \ref{fig:hardness} with a hardness ratio of $0.48$ and a 0.7--8.0~keV flux of $3\mathrm{\times10^{-12}erg~s^{-1}~cm^{-2}}$. We observe that the source shows clear spectral curvature in all observations where the flux is above $\sim 6\mathrm{\times10^{-12}erg~s^{-1}~cm^{-2}}$, though we cannot rule out curvature in the lower luminosity observations on the hard track which might occur above the energy range of the instruments used in this study. \\

\subsection{\textit{Chandra} Spectrum}

\begin{figure}
\centering

\includegraphics[width=.48\textwidth, angle=0]{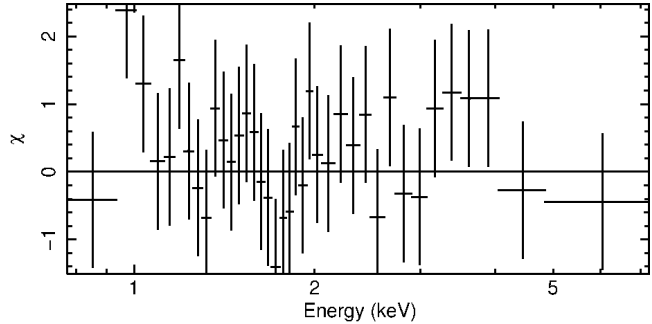}\\
\caption{IC 342 X-1 \textit{Chandra} energy spectrum residuals of the MCD fit.}
\label{fig:spec}
\end{figure}

The fit parameters of CH for a simple powerlaw, disc model, and combination are shown in Table \ref{table:fit}. The spectrum is best described by a disc model, though we note that the powerlaw model cannot be entirely excluded as it is only rejected with 2$\sigma$ confidence. Although X-1 has generally been observed in a hard, powerlaw-dominant state, the most recent \textit{Chandra} observation confirms that X-1 does undergo spectral transitions to an apparently disc-dominant state. The MCD model residuals for CH are shown in Figure \ref{fig:spec}.

\begin{table*}
\caption{\textit{Chandra} Spectral Fit Parameters}
\centering
%\centering

%\tabcolsep=0.1cm

\begin{tabular}{ccccccccccccc}  %{l*{10}{c}r}

\hline
Model & $N_\mathrm{H}$  & $\Gamma$ & $N_{pow}$ & $T_{\mathrm{in}}$ & $N_{disc}$ & $F_{X}$  & $Int. L_{X}$ & Disc Frac. & $\chi^2$(DoF) \\

\hline
(1) & (2) & (3) & (4) & (5) & (6) & (7) & (8) & (9) & (10)  \\
\hline

Pow &  $1.10^{+0.10}_{-0.09}$ & $2.98^{+0.15}_{-0.14}$ & $2.5^{+0.4}_{-0.4}$  & ... & ... &  $2.05^{+0.11}_{-0.11}$ &  $13.6^{+2.2}_{-1.7}$  &...& 138.70(108)  \\

\textbf{ MCD} & $0.56^{+0.06}_{-0.05}$ & ...  & ... & $0.91^{+0.05}_{-0.05}$ &  $0.23^{+0.06}_{-0.05}$ & $1.93^{+0.09}_{-0.09}$   & $6^{+4}_{-2}$ & ...& \textbf{98.99(108)}  \\

Pow+MCD & $0.67^{+0.06}_{-0.05}$ & $2.5$\# & $0.303$\# & $0.87^{+0.06}_{-0.06}$  & $0.24^{+0.09}_{-0.06}$ &  $1.99^{+0.08}_{-0.08}$ &  $6.4^{+0.6}_{-0.5}$ & \textgreater$0.71$ & 101.61(108)  \\

\hline
\end{tabular}
\label{table:fit}
%\end{center}
%

\begin{flushleft}
(1) Model used for spectral fit; (2) Total absorption column ($10^{22}\ \mathrm{cm^{-2}}$); (3) Power-law photon index; (4) Power-law normalization ($\times10^{-3}$); (5) Inner-disc temperature ; (6) Disc Normalization; (7) Absorbed flux (0.5--10 keV) ($\times 10^{-12}\ \mathrm{erg\ s^{-1}\ cm^{-2}}$); (8) Unabsorbed luminosity (0.5--10 keV) for $D=3.9\ \mathrm{Mpc}$ ($\times 10^{39}\ \mathrm{erg\ s^{-1}}$); (9) Disc fraction: intrinsic disc flux/total flux in 0.5--10~keV range; (10) $\chi^2$ and degrees of freedom.\\
\# denotes parameter held fixed during fitting\\
\end{flushleft}

\end{table*}

The most recent \textit{Chandra} observation is well fit by a single component MCD model. When fit with a two component MCD plus powerlaw model, the powerlaw component contributes less than 30~per~cent of the total flux in the 0.5--10~keV range. 

\section{Radio Observations} 
Radio jets have been observed from stellar mass BHBs, supermassive BHs and, recently, ULXs \citep{Webb12, Middleton13, Cseh14}. We carried out radio observations of X-1 using the Jansky Very Large Array (\textit{JVLA}) of the National Radio Astronomy Observatory (NRAO\footnote{The National Radio Astronomy Observatory is a facility of the National Science Foundation operated under cooperative agreement by Associated Universities, Inc.}) simultaneously with the \textit{Chandra} X-ray observations on 29 October, 2012 from 06:38 to 09:38. The longest baseline, A-array configuration, was used (program SD0164) to test the compactness of previously reported radio emission coincident with the ULX. We observed the source for 55~min at C-band ($4.5-6.5$~GHz) and X-band ($8.0-10.0$~GHz) both with $\sim$2~GHz bandwidth. The flux and bandpass calibrator referenced was 3C147, and the phase calibrator was J0228+6721. We processed our continuum data using the Common Astronomy Software Application package (\textsc{CASA}\footnote{http://casa.nrao.edu/}) version 4.0 using standard calibration procedures. We also used standard procedures for imaging, and used Briggs weighting with the robust parameter set to 0 to create radio images at both frequencies. The parameters of the resulting images, including the restoration beam size returned by the \textsc{clean} utility and the rms noise level are listed in Table \ref{table:radio}. 

\subsection{Search for Compact Core Emission}

\begin{figure}

\includegraphics[width=.45\textwidth]{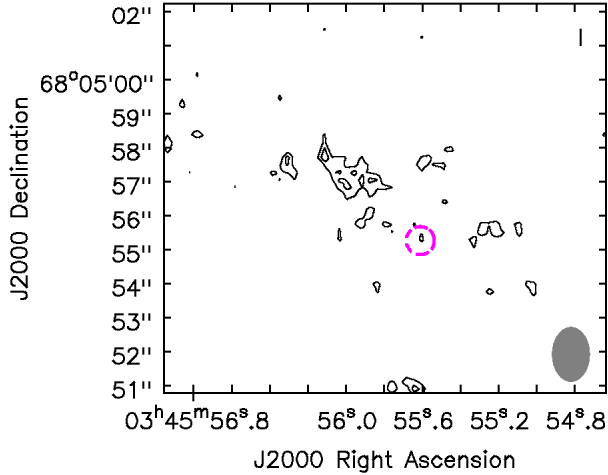}

\caption{5.5~GHz image from our most recent \textit{JVLA} observations. The image was produced with natural (Briggs=2) weighting where we have used a restoring beam size of 1.6"$\times$1.1" in the \textsc{clean} process to match the beam size of the previous \textit{VLA} observations where possible unresolved emission was reported \citet{Cseh12}. Contours are drawn at 2 and 3 times the rms level of $5.1~\mu\rm Jy~beam^{-1}$. The ULX error circle is indicated by a pink circle. No emission above the $3\sigma$ level is detected within the ULX region. The restoring beam is drawn as a filled ellipse in the bottom right corner.}
\label{fig:cbandb}

\end{figure}

We estimated the maximum radio flux from X-1 at the position determined by \citet{Feng09} of R.A. = $3^{h}45^{m}55.612^{s}$, decl. = $+68^{\circ}04'55.29''$ (J2000.0) with a $0.21''$ error circle at the 90~per~cent confidence level. Using a conservative search region with twice the error circle radius, we find no radio emission above the $3\sigma$ rms noise level at either frequency. We can therefore place upper limits on the compact core emission from the new \textit{JVLA} observations at $14~\mu$Jy at 5.5~GHz. To bring out emission from the surrounding nebula, we additionally imaged the 5.5~GHz data using robust = 2 (natural) weighting, but still find no significant emission associated with the position of the ULX. Using a restoring beam size in the \textsc{clean} task of the previous \textit{VLA} observations, $1.6''\times 1.1''$, we are able to bring out some of the features of the surrounding radio nebula but still detect no significant emission in the ULX region (see Figure \ref{fig:cbandb}). This non-detection agrees with observations with the \textit{EVN} reported by \citet{Cseh12} and supports their conclusion that the emission reported in the 2007 and 2008 \textit{VLA} observations was likely a knot of radio emission from the surrounding nebula which is over-resolved in the newest observations because of their larger array configuration.

To investigate whether the previously reported emission was due to a knot of emission or alternatively a transient event, we reanalysed the previous \textit{VLA} data consisting of two 3.2 hour observations at 4.8~GHz. One observation was made using the VLA's B-array on 6, December 2007 and a second with the C-array on 25, April 2008, 141 days later. We posit that in combining the data from the two arrays,  extended emission from the radio nebula observed by the C-array could appear artificially compacted. To test this, we combined the B and C array data and compared the peak emission in the ULX region when the uv range (the range of included baseline lengths) is restricted to be greater than $35~\rm k\lambda$, thus excluding extended emission from the shortest baselines, while still including data from the longest baselines of the C-array. If a point source is present, it should be visible to all of the long baselines and the flux of the compact emission should not decrease significantly when the short baselines are omitted. Using Briggs = 0 weighting for both cases (the original combined versus the uv-restricted combined data), we find that the maximum flux in the ULX region decreases from 91$~\mu$Jy (6 times the rms level of 15$~\mu\rm Jy~beam^{-1}$) to 63$~\mu$Jy (3 times the rms level of 21$~\mu\rm Jy~beam^{-1}$) when the short baselines are omitted. Thus we do not find a highly significant detection using conditions optimized for compact emission, in agreement with the discussion of \citet{Cseh12}.

\subsection{Diffuse Emission Near X-1}

\begin{figure}

\includegraphics[width=.45\textwidth]{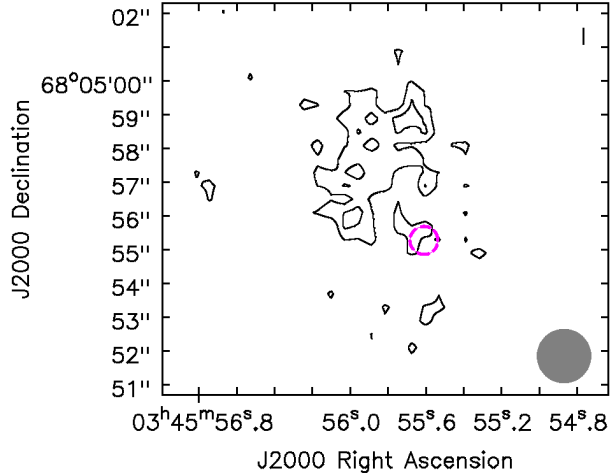}

\caption{Previous 2011 A-array 5~GHz image with natural (Briggs=2) weighting where we have used a restoring beam of 1.6"$\times$1.6" to bring out the extended features in the surrounding nebula. Contours are drawn at 2 and 3 times the rms level of $13.3~\mu\rm Jy~beam^{-1}$. The ULX error circle is indicated by a pink circle. The maximum level of emission within the ULX region is 2.4 times the rms level. The restoring beam is drawn as a filled ellipse in the bottom right corner.}
\label{fig:millercbandb}

\end{figure}

The emission near X-1 could be attributable to a knot from the surrounding radio nebula, or possibly to expanded ejecta from the ULX as was observed in the ULX Holmberg II X-1 by \citet{Cseh14}. To estimate the minimum size of a the knot near the ULX, we analysed 4.9~GHz observations that we made with the \textit{JVLA} in the A-array on 3 July, 2011 (project 11A-173). The source was observed for 97 minutes. The flux and bandpass calibrator referenced was 3C147, and the phase calibrator was J0410+7656. These observations were not coincident with X-ray observations, and were taken in the early testing mode of the new correlator and restricted to a smaller bandwidth of 256~MHz and centred on 4.9~GHz and are subsequently at a slightly lower resolution to our most recent \textit{JVLA} data with a naturally weighted beam size of $0.54'' \times 0.45''$. Again, we find no significant emission in the error circle of the ULX with a peak flux in the ULX region of 1.9 times the rms level of $8.9~\mu\rm Jy~beam^{-1}$. Creating an image using a restoring beam size of $1.6''\times 1.6''$ in the \textsc{clean} task, we are able to bring out some of the extended emission to the NE of the ULX, shown in Figure \ref{fig:millercbandb}, though the emission in the ULX region is still only 2.4 times the rms level of $13.3~\mu$Jy. The $0.5''$ beam diameter of the 2011 \textit{JVLA} image is 1/7 times that of the previous \textit{VLA} observation with beam size 1.6"$\times$1.1." In order for the $\sim63~\mu$Jy source to fall below the $3\sigma$ rms noise level, it must be spread over at least two beams in the 2011 data, and thus larger than $\sqrt{2} \times 0.5''$ or $\gtrsim 13$~pc at 3.9~Mpc. This minimum is comparable to size of ejecta observed in Holmberg II X-1 \citep{Cseh14}. Therefore, from our current data we cannot rule out the possibility of emission due to expanding ejecta from the ULX, though further observations would be necessary to study such an interpretation.

\begin{table}

\caption{Radio Image Parameters and Source Detection Limits}
\centering
\tabcolsep=0.15cm
%\centering
%\tiny
\begin{tabular}{ccccccc}

\hline
Date & Config. & Freq. & Width & R & $3~\sigma$ & $Beam size$\\
\hline
(1) & (2) & (3) & (4) & (5) & (6) & (7) \\ 
\hline
10/29/12 & A & 9 & 2 & 0 & 18.9 & 0.21 $\times$ 0.14 \\
10/29/12 & A & 5.5 & 2 & 0 & 19.2 & 0.31 $\times$ 0.23 \\
10/29/12 & A & 5.5 & 2 & 2 & 13.5 & 0.46 $\times$ 0.37 \\

07/03/11 & A & 4.9 & 0.256 & 2 & 26.7  & 0.54 $\times$ 0.45 \\

04/25/08 & C & \multirow{2}{*}{4.8} & \multirow{2}{*}{0.05} & \multirow{2}{*}{0} & \multirow{2}{*}{45} &  \multirow{2}{*}{1.6 $\times$ 1.2 }\\

12/06/07 & B & &  &  &  &  \\

\hline
\end{tabular}
\label{table:radio}
\begin{flushleft}
\small{(1) Observation date; (2) \textit{VLA} array configuration; (3) Central Frequency of Bandwidth (GHz) ; (4) Bandwidth (GHz); (5) Briggs weighting parameter used to make image ; (6) Three times the rms value ($\mu\rm Jy~beam^{-1}$); (7) Beam size (arcsec)$^{2}$ }
\end{flushleft}

\end{table}

\section{Discussion}

\begin{figure}

\begin{minipage}{1\linewidth}
\centering
\includegraphics[width=.7\linewidth, angle=270]{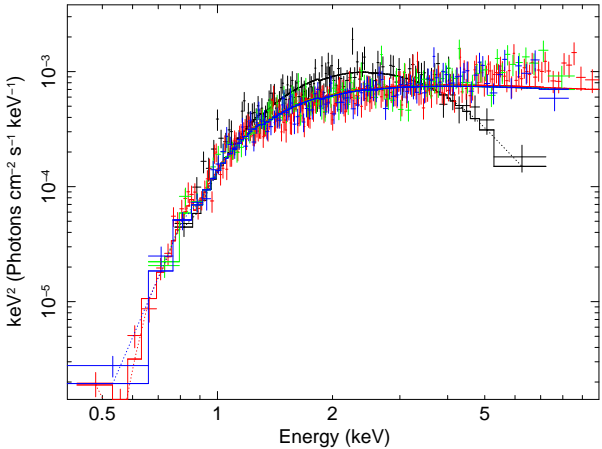}\\
\caption{Unfolded model and spectrum of XMM3 (RGB) and CH (black) observations plotted in $\mathrm{E^2f(E)}$.}
\label{fig:chxmm}
\end{minipage}
\end{figure}

%HID
From the hardness-intensity diagram in Figure \ref{fig:hardness}, it is clear that IC 342 X-1 (X-1) is in a different spectral state in the most recent \textit{Chandra} observations than has been observed previously. However, the flux of X-1 in this state is the same as in harder observations, therefore the spectral state of X-1 cannot be determined by the luminosity alone. This is illustrated in Figure \ref{fig:chxmm}, where the model and spectra are compared for the recent \textit{Chandra} observation and a previous \textit{XMM-Newton} observation of the source. These observations are very similar in flux (5 per cent difference using best-fit single component models), but have very different spectral shapes. The spectral states of BHBs \citep{Remillard06} and some ULXs have also been shown to be degenerate with luminosity, and \citet{Grise10} showed similarly for the ULX Ho II X-1 that spectral transitions occurred without a significant change in luminosity.

%HID

%mass
Based on high-energy spectral curvature, X-1 appears to be in a super-Eddington state in its highest luminosity observations, as has been suggested by previous authors \citep{Watarai01, Ebisawa03, Feng09, Yoshida12}. This spectral curvature has been theorized to be the result of heavy Comptonization by a cool, optically thick corona when the source is in a high accretion state \citep{Gladstone09}. However, at low luminosity X-1 appears to transition to a sub or near-Eddington accretion state where high energy curvature is less significant. \citet{Sutton13} suggested that many ULXs are found in a `broadened' disc state at their lowest luminosities, and transition into `hard ultraluminous' and `soft ultraluminous' states with increasing luminosity, which may represent a transition from near-Eddington to super-Eddington accretion. The new \textit{Chandra} observation is consistent with the broadened disc classification. However, we note that there is no path to a classical disc state in the \citet{Sutton13} classification scheme. 

The ASCA1 and ASCA2 observations have been interpreted as a transition between a disc-dominated state and a powerlaw-dominated state \citep{Kubota01}. The ASCA1 observation, which is well modeled by a disc spectrum, appears on the hard track of the observations in our sample and its hardness is not significantly different from the ASCA2 observation. \citet{Sutton13} point out that a hard ultraluminous spectrum with some soft excess could be misidentified as a broadened disc spectrum, which could be an explanation for the the ASCA1 observation.

Because the \textit{Chandra} spectrum appears consistent with disc-dominated emission, we use the fitted disc properties to estimate the mass of the BH based on a Shakura-Sunyaev disc. The bolometric luminosity is related to the BH mass and disc temperature as \citep{Makishima00},\\

\begin{equation}
 L_{\rm bol} = 7.2\times10^{38}\left(\frac{\zeta}{0.41}\right)^{-2} \left(\frac{\kappa}{1.7}\right)^{-4} \alpha^{2}\left(\frac{M}{10~\mathrm{M_{\odot}}}\right)^{2} \left(\frac{ T_{\rm in}}{{\rm keV}}\right)^{4}
\end{equation}\\
where $\zeta \sim 0.41$ is a factor to correct for the small difference between the radius where the maximum temperature occurs and the true $R_{\rm in}$, $\kappa \sim 1.7$ is a spectral hardening factor to correct the measured disc blackbody temperature to the effective inner-disc temperature. $\alpha$ depends on the black hole's spin and is equal to 1 for a non-spinning Schwarzchild black hole and $\frac{1}{6}$ for a maximally rotating Kerr black hole. $L_{\rm bol} = 2\pi D^{2} f_{\rm bol}/\mathrm{cos}i$, where $i$ is the inclination of the system. We approximate the bolometric flux as the MCD model flux integrated over 0.01--100~keV and assume the canonical values for $\zeta$ and $\kappa$.

The mass of the BH is then given by,\\

\begin{equation}
M=10\left(\frac{ T_{\rm in}}{\rm{keV}}\right)^{-2}~\alpha^{-1}~\left( \frac{L_{\rm bol}}{7.2\times10^{38}\mathrm{\rm{erg\ s}^{-1}}}\right)^{1/2}
\end{equation}\\
For kT~=~0.91~keV and $L_{\rm bol} = 6\times10^{39}{~\rm erg\ s}^{-1}~/$cos$i$ and taking the limits of $\alpha=$1 and 1/6, this yields a mass range of $30~\mathrm{M_{\odot}} \lesssim M\sqrt{\mathrm{cos}i}\lesssim 200~\mathrm{M_{\odot}}$.\\

At high accretion rates, the disc is expected to be described by an optically thick `slim' disc state, in which cooling by advection becomes a non-negligible factor. As modelled by \citet{Watarai01}, the luminosity-temperature relationship in the slim disc regimes deviates from the standard $L\propto T^{4}$ relationship such that, for a super-Eddington system, the mass calculated at a given luminosity and temperature for a thin disc would yield an over-estimate. Because X-1 may be near-Eddington, we additionally fit the \textit{Chandra} spectrum with a slim disc model developed by \citet{Kawaguchi03}. When fitting with a slim disc model, some authors have found disc dominated ULX spectra to be better fit by a two-component model that includes Comptonization in addition to a slim disc component rather than the slim disc alone \citep[e.g.][]{Middleton11, Middleton12}. We find that the \textit{Chandra} spectrum is well fit by the \citet{Kawaguchi03} model number 7 which includes a Comptonized local spectrum and relativistic effects. This model has also been used to fit other ULX spectra \citep[e.g.][]{Foschini06, Vierdayanti06, Godet12, Bachetti13}. The parameters of the slim disc fit are listed in Table \ref{table:slimfit}. This model assumes a non-rotating BH and that the source is face-on, due to complications of the model for inclined geometries. We find a mass of $77^{+27}_{-10}~\mathrm{M_{\odot}}$ with the slim disc model, which would increase in the case of a rotating black hole. The best-fit mass accretion rate is 9.7 in units of $L_{\rm Edd}/c^2$, where $\dot{M} = L/\eta c^2$. The Eddington fraction is $\dot{M}/ M_{\rm Edd} = \eta \dot{M}$, where the efficiency factor, $\eta $, depends on the accretion rate and mass of the BH. \citet{Ebisawa03} plot the values of $\eta $ for a slim disc model as a function of mass and $\dot{M}$ in their Figure 7. For our best-fit slim disc model parameters, $\eta \sim 1/16$,  which yields an accretion rate of $\sim 60$ per cent of the Eddington rate.

\begin{table}
\caption{Slim Disc Fit Parameters}

\tabcolsep=0.1cm

\begin{tabular}{cccccc} 

\hline
$N_\mathrm{H}$ & M & \.{M}  &$\alpha$ & $N_{slim}$  & $\chi^2$(DoF) \\

\hline
(1) & (2) & (3) & (4) & (5) & (6)   \\
\hline

 $0.71^{+0.07}_{-0.06}$ & $77^{+27}_{-9}$ & $9.7^{+0.8}_{-2.1}$ & $0.10^{+0.03}_{-0.04}$ &  $6.56\times 10^{-6}$\# & 99.66(107) \\

\hline

\end{tabular}
\label{table:slimfit}

\begin{flushleft}
(1) Total absorption column ($10^{22}\ \mathrm{cm^{-2}}$); (2) Black hole mass; (3) Mass accretion rate ($\mathrm{L_{Edd}}/c^2$); (4) Viscosity; (5) Normalization ($(10~\rm kpc/d)^2$); (6) $\chi^2$ and degrees of freedom.\\
\# denotes parameter held fixed during fitting
\end{flushleft}

\end{table}

In \textit{JVLA} observations simultaneous with the \textit{Chandra} observations, no coincident emission is detected above the 3$\sigma$ noise level of 13.5$~\mu$Jy, which is consistent with X-1 being in a thermal X-ray state and the derived BH mass. This non-detection is also consistent with 1.6~GHz \textit{EVN} observations in June 2011 when the source was observed to be in an apparent hard state by \textit{Swift} \citep{Cseh12} and with observations with the \textit{JVLA} in 2011. We therefore cannot independently confirm the thermal state with the current radio observations, which do not show radio emission being quenched compared to other spectral states \citep{Corbel00}. Additionally, if X-1 is near the low end of the estimated mass range, the radio emission in the hard state would likely be a few $~\mu$Jy \citep{Merloni03}, below the sensitivity of current observations.

After reanalysis of 2007 and 2008 VLA observations, we agree with the conclusion of \citet{Cseh12} that the previously reported emission was likely not due to compact emission from a radio flare, though we cannot entirely rule out a radio flare with the current data. Microquasar flares \citep{Blundell11, Corbel12, Punsly13} are observed to be approximately $1-10$~Jy at a distance of $\sim8$~pc, which corresponds to $4-40~\mu$Jy when scaled for the distance of 3.9~Mpc to IC 342, on the order of the unresolved emission of 63$~\mu$Jy we find in the 2007-2008 VLA data. 

If the emission in the previous VLA observations was due to a radio knot, we can set a lower limit to its linear size based on the resolution of the 2011 \textit{JVLA} observations. The beam diameter of the naturally weighted 4.9~GHz image is 0.5", and in order for the source observed in previous VLA observations to fall below $3\sigma$ per resolution element, it must be spread across more than two beams in the higher resolution observations, corresponding to a linear size of $\gtrsim 13$~pc at 3.9~Mpc. This is larger but still comparable to size of ejecta observed near the ULX Holmberg II X-1 by \citet{Cseh14}, and one possible interpretation is that the observed knot of emission is expanded ejecta from the ULX. However, the radio field surrounding X-1 is complex and higher signal to noise observations would be necessary to make any strong conclusions. 
%radio

\section{Acknowledgements}
Support for this work was provided by the National Aeronautics and Space Administration through \textit{Chandra} Award Number GO2--13036X issued by the \textit{Chandra} X-ray Observatory Center, which is operated by the Smithsonian Astrophysical Observatory for and on behalf of the National Aeronautics Space Administration under contract NAS8--03060. St{\'e}phane Corbel acknowledges funding support from the French Research National Agency: CHAOS project ANR-12-BS05-0009\footnote{http://www.chaos-project.fr} and financial support from the UnivEarthS Labex program of Sorbonne Paris Cit\'{e} (ANR-10-LABX-0023 and ANR-11-IDEX-0005-02).). We thank the anonymous referee for her/his comments and suggestions which contributed to the improvement of the paper.


\begin{thebibliography}{}
\setlength{\itemsep}{-0.0em}
\setlength{\parsep}{4pt}
\setlength{\baselineskip}{13pt}

%\bibitem[Abramowicz et al.(1988)]{Abramowicz88} Abramowicz, M.~A., Czerny, B., Lasota, J.~P., \& Szuszkiewicz, E.\ 1988, \apj, 332, 646 

\bibitem[Arnaud(1996)]{Arnaud96} Arnaud, K.~A.\ 1996, 
Astronomical Data Analysis Software and Systems V, 101, 17 

\bibitem[Bachetti et al.(2013)]{Bachetti13} Bachetti, M., Rana, 
V., Walton, D.~J., et al.\ 2013, \apj, 778, 163 

\bibitem[Begelman(2002)]{Begelman02} Begelman, M.~C.\ 2002, \apjl, 
568, L97

%\bibitem[Belloni(2010)]{Belloni10} Belloni, T.~M.\ 2010, Lecture Notes in Physics, Berlin Springer Verlag, 794, 53 

%\bibitem[Belloni et al.(2005)]{Belloni05} Belloni, T., Homan, J., Casella, P., et al.\ 2005, \aap, 440, 207 

\bibitem[Blundell et al.(2011)]{Blundell11} Blundell, K.~M., Schmidtobreick, L., \& Trushkin, S.\ 2011, \mnras, 417, 2401 

\bibitem[Colbert 
\& Mushotzky(1999)]{Colbert99} Colbert, E.~J.~M., \& Mushotzky, R.~F.\ 1999, \apj, 519, 89 

\bibitem[Cseh et al.(2011)]{Cseh11} Cseh, D., Lang, C., 
Corbel, S., Kaaret, P., \& Gris{\'e}, F.\ 2011, IAU Symposium, 275, 325 

\bibitem[Corbel et 
al.(2000)]{Corbel00} Corbel, S., Fender, R.~P., Tzioumis, A.~K., et al.\ 2000, \aap, 359, 251

\bibitem[Corbel et al.(2012)]{Corbel12} Corbel, S., Dubus, G., 
Tomsick, J.~A., et al.\ 2012, \mnras, 421, 2947 

\bibitem[Cseh et al.(2012)]{Cseh12} Cseh, D., Corbel, S., 
Kaaret, P., et al.\ 2012, \apj, 749, 17 

\bibitem[Cseh et al.(2014)]{Cseh14} Cseh, D., Kaaret, P., 
Corbel, S., et al.\ 2014, \mnras, 439, L1 


\bibitem[Ebisawa et al.(2003)]{Ebisawa03} Ebisawa, K., {\.Z}ycki, 
P., Kubota, A., Mizuno, T., \& Watarai, K.-y.\ 2003, \apj, 597, 780

 

\bibitem[Falcke et 
al.(2004)]{Falcke04} Falcke, H., K{\"o}rding, E., \& Markoff, S.\ 2004, \aap, 414, 895 

%\bibitem[Fender et al.(1999)]{Fender99} Fender, R., Corbel, S., Tzioumis, T., et al.\ 1999, \apjl, 519, L165 
%\bibitem[Fender et al.(2004)]{Fender04} Fender, R.~P., Belloni, T.~M., \& Gallo, E.\ 2004, \mnras, 355, 1105 

%\bibitem[Feng \& Kaaret(2006)]{Feng06} Feng, H., \& Kaaret, P.\ 2006, \apjl, 650, L75

\bibitem[Feng 
\& Kaaret(2008)]{Feng08} Feng, H., \& Kaaret, P.\ 2008, \apj, 675, 1067

\bibitem[Feng 
\& Kaaret(2009)]{Feng09} Feng, H., \& Kaaret, P.\ 2009, \apj, 696, 1712  %xmm obs

%\bibitem[Feng 
%\& Soria(2011)]{Feng11} Feng, H., \& Soria, R.\ 2011, New Astronomy Reviews, 55, 166 

%\bibitem[Frank et al.(2002)]{Frank02} Frank, J., King, A., \& Raine, D.~J.\ 2002, Accretion Power in Astrophysics, by Juhan Frank and Andrew King and Derek Raine, pp.~398.~ISBN 0521620538.~Cambridge, UK: Cambridge University Press, February 2002., 

\bibitem[Foschini et al.(2006)]{Foschini06} Foschini, L., Ebisawa, 
K., Kawaguchi, T., et al.\ 2006, Advances in Space Research, 38, 1378 

\bibitem[Gladstone et al.(2009)]{Gladstone09} Gladstone, J.~C., 
Roberts, T.~P., \& Done, C.\ 2009, \mnras, 397, 1836

\bibitem[Godet et al.(2012)]{Godet12} Godet, O., Plazolles, B., 
Kawaguchi, T., et al.\ 2012, \apj, 752, 34 


%\bibitem[Gon{\c c}alves \& Soria(2006)]{Gon06} Gon{\c c}alves, A.~C., \& Soria, R.\ 2006, \mnras, 371, 673 


\bibitem[Gris{\'e} et al.(2010)]{Grise10} Gris{\'e}, F., 
Kaaret, P., Feng, H., Kajava, J.~J.~E., 
\& Farrell, S.~A.\ 2010, \apjl, 724, L148 

%\bibitem[Harrison et al.(2013)]{Harrison13} Harrison, F.~A., Craig, W.~W., Christensen, F.~E., et al.\ 2013, \apj, 770, 103

%\bibitem[Ishida et al.(2011)]{Ishida11} Ishida, M., Tsujimoto, M., Kohmura, T., et al.\ 2011, \pasj, 63, 657 


\bibitem[Kaaret et al.(2001)]{Kaaret01} Kaaret, P., Prestwich, 
A.~H., Zezas, A., et al.\ 2001, \mnras, 321, L29 

%\bibitem[Kaaret et al.(2003)]{Kaaret03} Kaaret, P., Corbel, S., 
%Prestwich, A.~H., \& Zezas, A.\ 2003, Science, 299, 365

\bibitem[Kaaret et al.(2004)]{Kaaret04} Kaaret, P., Ward, M.~J., 
\& Zezas, A.\ 2004, \mnras, 351, L83 

\bibitem[Kaaret 
\& Corbel(2009)]{Kaaret09a} Kaaret, P., \& Corbel, S.\ 2009, \apj, 697, 950 

%\bibitem[Kaaret et al.(2009)]{Kaaret09} Kaaret, P., Feng, H., \& Gorski, M.\ 2009, \apj, 692, 653 


%\bibitem[Kaaret 
%\& Corbel(2009)]{Kaaret09b} Kaaret, P., \& Corbel, S.\ 2009, \apj, 697, 950 

\bibitem[Kajava 
\& Poutanen(2009)]{Kajava09} Kajava, J.~J.~E., \& Poutanen, J.\ 2009, \mnras, 398, 1450 

%\bibitem[Kalberla et 
%al.(2005)]{Kalberla05} Kalberla, P.~M.~W., Burton, W.~B., Hartmann, D., et al.\ 2005, \aap, 440, 775 

\bibitem[Kawaguchi(2003)]{Kawaguchi03} Kawaguchi, T.\ 2003, \apj, 
593, 69 

%\bibitem[King 
%\& Pounds(2003)]{King03} King, A.~R., \& Pounds, K.~A.\ 2003, \mnras, 345, 65


\bibitem[King(2009)]{King09} King, A.~R.\ 2009, \mnras, 393, 
\bibitem[Kong(2003)]{Kong03} Kong, A.~K.~H.\ 2003, \mnras, 
346, 265 


\bibitem[Kubota et al.(2001)]{Kubota01} Kubota, A., Mizuno, T., 
Makishima, K., et al.\ 2001, \apjl, 547, L119 
%\bibitem[Kubota et al.(2001)]{2001ApJ...547L.119K} Kubota, A., Mizuno, T., 
%Makishima, K., et al.\ 2001, \apjl, 547, L119 
\bibitem[Kubota 
\& Makishima(2004)]{Kubota04} Kubota, A., \& Makishima, K.\ 2004, \apj, 601, 428 

\bibitem[Mak et al.(2008)]{Mak08} Mak, D.~S.~Y., Pun, 
C.~S.~J., \& Kong, A.~K.~H.\ 2008, \apj, 686, 995 

\bibitem[Mak et al.(2011)]{Mak11} Mak, D.~S.~Y., Pun, 
C.~S.~J., \& Kong, A.~K.~H.\ 2011, \apj, 728, 10 

\bibitem[Makishima et al.(2000)]{Makishima00} Makishima, K., 
Kubota, A., Mizuno, T., et al.\ 2000, \apj, 535, 632 

%\bibitem[Markwardt(2009)]{Markwardt09} Markwardt, C.~B.\ 2009, Astronomical Data Analysis Software and Systems XVIII, 411, 251 

%\bibitem[McClintock et al.(2006)]{McClintock06} McClintock, J.~E., Shafee, R., Narayan, R., et al.\ 2006, \apj, 652, 518 

\bibitem[Merloni et al.(2003)]{Merloni03} Merloni, A., Heinz, S., 
\& di Matteo, T.\ 2003, \mnras, 345, 1057 

\bibitem[Miyawaki et al.(2009)]{Miyawaki09} Miyawaki, R., 
Makishima, K., Yamada, S., et al.\ 2009, \pasj, 61, 263 

\bibitem[Middleton et al.(2011)]{Middleton11} Middleton, M.~J., 
Sutton, A.~D., \& Roberts, T.~P.\ 2011, \mnras, 417, 464

\bibitem[Middleton et al.(2012)]{Middleton12} Middleton, M.~J., 
Sutton, A.~D., Roberts, T.~P., Jackson, F.~E., 
\& Done, C.\ 2012, \mnras, 420, 2969 

\bibitem[Middleton et al.(2013)]{Middleton13} Middleton, M.~J., 
Miller-Jones, J.~C.~A., Markoff, S., et al.\ 2013, \nat, 493, 187 

%\bibitem[Miller et al.(2003)]{Miller03} Miller, J.~M., Fabbiano, 
%G., Miller, M.~C., \& Fabian, A.~C.\ 2003, \apjl, 585, L37 

\bibitem[Mizuno et al.(2001)]{Mizuno01} Mizuno, T., Kubota, A., 
\& Makishima, K.\ 2001, \apj, 554, 1282 

\bibitem[Okada et al.(1998)]{Okada98} Okada, K., Dotani, T., 
Makishima, K., Mitsuda, K., \& Mihara, T.\ 1998, \pasj, 50, 25 

\bibitem[Pakull 
\& Mirioni(2002)]{Pakull02} Pakull, M.~W., \& Mirioni, L.\ 2002, arXiv:astro-ph/0202488 


\bibitem[Pintore et al.(2014)]{Pintore14} Pintore, F., Zampieri, 
L., Wolter, A., \& Belloni, T.\ 2014, arXiv:1401.6815 

\bibitem[Poutanen et al.(2007)]{Poutanen07} Poutanen, J., 
Lipunova, G., Fabrika, S., Butkevich, A.~G., 
\& Abolmasov, P.\ 2007, \mnras, 377, 1187 

\bibitem[Punsly 
\& Rodriguez(2013)]{Punsly13} Punsly, B., \& Rodriguez, J.\ 2013, \mnras, 2127 

\bibitem[Rana et al.(2014)]{Rana14} Rana, V., Harrison, F.~A., 
Bachetti, M., et al.\ 2014, arXiv:1401.4637 

\bibitem[Remillard 
\& McClintock(2006)]{Remillard06} Remillard, R.~A., \& McClintock, J.~E.\ 2006, \araa, 44, 49 

\bibitem[Roberts et al.(2003)]{Roberts03} Roberts, T.~P., Goad, 
M.~R., Ward, M.~J., \& Warwick, R.~S.\ 2003, \mnras, 342, 709 

\bibitem[Roberts et al.(2004)]{Roberts04} Roberts, T.~P., 
Warwick, R.~S., Ward, M.~J., \& Goad, M.~R.\ 2004, \mnras, 349, 1193 %chandra obs

%\bibitem[Shakura \& Sunyaev(1973)]{Shakura73} Shakura, N.~I., \& Sunyaev, R.~A.\ 1973, \aap, 24, 337

\bibitem[Stobbart et al.(2006)]{Stobbart06} Stobbart, A.-M., 
Roberts, T.~P., \& Wilms, J.\ 2006, \mnras, 368, 397 

\bibitem[Sutton et al.(2013)]{Sutton13} Sutton, A.~D., Roberts, 
T.~P., \& Middleton, M.~J.\ 2013, \mnras, 2087

%\bibitem[Terlouw \& Vogelaar(2013)]{KapteynPackage} J.~P. {Terlouw} and M.~G.~R. {Vogelaar}.  {Kapteyn Package, version 2.2.1b16}. Kapteyn Astronomical Institute, Groningen, September 2013.\\ Available from http://www.astro.rug.nl/software/kapteyn/.

\bibitem[Tikhonov \& Galazutdinova(2010)]{Tikhonov10} Tikhonov, N.~A., \& Galazutdinova, O.~A.\ 2010, Astronomy Letters, 36, 167 

%\bibitem[Titarchuk \& Lyubarskij(1995)]{Titarchuk95} Titarchuk, L., \& Lyubarskij, Y.\ 1995, \apj, 450, 876 

\bibitem[Vierdayanti et al.(2006)]{Vierdayanti06} Vierdayanti, K., 
Mineshige, S., Ebisawa, K., \& Kawaguchi, T.\ 2006, \pasj, 58, 915 

\bibitem[Vierdayanti et al.(2010)]{Vierdayanti10} Vierdayanti, K., 
Done, C., Roberts, T.~P., \& Mineshige, S.\ 2010, \mnras, 403, 1206 

\bibitem[Wang et al.(2004)]{Wang04} Wang, Q.~D., Yao, Y., 
Fukui, W., Zhang, S.~N., \& Williams, R.\ 2004, \apj, 609, 113

\bibitem[Watarai et al.(2001)]{Watarai01} Watarai, K.-y., Mizuno, 
T., \& Mineshige, S.\ 2001, \apjl, 549, L77 

\bibitem[Webb et al.(2012)]{Webb12} Webb, N., Cseh, D., Lenc, 
E., et al.\ 2012, Science, 337, 554 

\bibitem[Yoshida et al.(2012)]{Yoshida12} Yoshida, T., Isobe, N., 
Mineshige, S., et al.\ 2012, arXiv:1212.0994 %Two power law states

\end{thebibliography}
\end{document}